# Exploring Hydride Formation in Stainless Steel Revisits Theory of Hydrogen Embrittlement

Cem Örnek[1*], Alfred Larsson[2], Mubashir Mansoor[1,3], Fan Zhang[4], Gary S Harlow[5,6], Robin Kroll[7], Francesco Carlà[8], Hadeel Hussain[8], Bora C Derin[1], Ulf Kivisäkk[9], Dirk L Engelberg[7], Edvin Lundgren[2], and Jinshan Pan[10*]

[1]Istanbul Technical University, Department of Metallurgical and Materials Engineering, Istanbul, Turkey
[2]Lund University, Division of Synchrotron Radiation Research, Lund, Sweden
[3]Istanbul Technical University, Department of Applied Physics, Istanbul, Turkey
[4]University of Sussex, Department of Engineering and Design, Brighton, United Kingdom
[5]Malmö University, Materials Science and Applied Mathematics, Malmö, Sweden
[6]MaxIV, Lund University, Lund, Sweden
[7]University of Manchester, Department of Materials, Manchester, United Kingdom
[8]Diamond Light Source, Didcot, United Kingdom
[9]Sandvik Materials Technology, Sandviken, Sweden
[10]KTH Royal Institute of Technology, Division of Surface and Corrosion Science, Stockholm, Sweden

*Corresponding author: Email: cornek@itu.edu.tr, tel.: +90 212 285 6864
*Corresponding author: Email: jinshanp@kth.se, tel.: +46 8 790 67 39
Email: alfred.larsson@sljus.lu.se, tel.: +46 46 222 42 21
Email: mansoor17@itu.edu.tr, tel.: +90 212 285 6864
Email: gary.harlow@mau.se, tel.: +46 46 222 71 33
Email: zhangfan@sussex.ac.uk , tel.: +44 770 4466957
Email: robin.kroll@postgrad.manchester.ac.uk, tel.: +44 161 306 2919
Email: francesco.carla@diamond.ac.uk, tel.: +44 1235 778023
Email: hadeel.hussain@diamond.ac.uk, tel.: +44 1235 778352
Email: bderin@itu.edu.tr, tel.: +90 212 285 3367
Email: ulf.kivisakk@sandvik.com, tel.: +46 26 26 49 30
Email: dirk.engelberg@manchester.ac.uk, tel.: +44 161 306 5952
Email: edvin.lundgren@sljus.lu.se, tel.: +46 46 222 42 21

## Abstract
Various mechanisms have been proposed for hydrogen embrittlement, but the causation of hydrogen-induced material degradation has remained unclear. This work shows hydrogen embrittlement due to phase instability (decomposition). In-situ diffraction measurements revealed metastable hydrides formed in stainless steel, typically declared as a non-hydride forming material. Hydride formation is possible by increasing the hydrogen chemical potential during electrochemical charging and low defect formation energy of hydrogen interstitials. Our findings demonstrate that hydrogen-induced material degradation can only be understood if measured in situ and in real-time during the embrittlement process.

## One Sentence Summary
The causation of hydrogen embrittlement is phase separation reactions.

# Introduction

Our modern society employs materials under increasingly harsh conditions than ever before. Technologically advanced components are often exposed to aggressive environments under high stresses, threatening the material's performance and lifetime. In applications where corrosion is the prime concern, metallic structures are often cathodically protected to prevent corrosion damage. However, cathodic protection is associated with the liberation of atomic hydrogen, which enters the material microstructure. Hydrogen in metals is typically detrimental, reducing the fracture strength, a phenomenon known as hydrogen embrittlement. Among all materials, high-strength steels are most susceptible to hydrogen-induced damage. Hydrogen down to ppb levels is enough to cause unexpected catastrophic failure [1]. Stainless steels, especially super austenitic and duplex grades, are workhorse materials in the oil and gas industry. However, due to hydrogen-induced cracks, some damages have been reported for cathodically protected duplex stainless steel in subsea applications [2, 3]. Hence, there are emergent risks for hydrogen embrittlement which have remained less understood.

Numerous research has been conducted to understand the hydrogen embrittlement mechanism, and various mechanisms have been proposed [1, 4-8]. In principle, hydrogen embrittlement occurs due to lattice decohesion or localized plasticity, and all other proposed mechanisms are derivatives of these events. For example, hydrogen immobilizes dislocation motion and causes a reduction in the binding energy between atoms resulting in decohesion, or it enhances dislocation activities inducing excessive localized plasticity [4]. In the end, the manifestation, in either case, is a material fracture. So, the proposed mechanisms point to the same outcome, which enflames whether there is a root mechanism that causes hydrogen embrittlement. This paper fosters our understanding of hydrogen embrittlement and demonstrates that the degradation can be caused by local decomposition due to phase instability (decomposition).

It is essential to summarise the knowledge we have gained from the rigorous research conducted about the hydrogen embrittlement of stainless steel. We showed that electrochemical hydrogen charging causes extensive lattice strains in duplex stainless steel microstructure [9-14], which can only be seen when measured in situ and in real time. The hydrogen-induced stresses concentrated primarily on the austenite phase [9-11], with near-surface grains deformed more heavily than bulk grains [12, 13]. In-situ time-resolved grazing-incidence x-ray diffraction (GIXRD) experiments during hydrogen charging showed diffraction peak splitting of the austenite phase, which gradually became more distinctive with increasing the magnitude and hydrogen absorption time [13]. The peaks existed only during the hydrogen charging process, indicating a metastable character [13]. We hypothesized the peak splitting as the evolution of metastable quasi-hydrides but have not provided a conclusive explanation for the formation mechanism [13], which we do herein.

Furthermore, we showed that the surface is most reactive to lattice strains and concluded that these metastable structures might play a vital role in super duplex stainless steel [9, 11-13]. A recent study showed that super duplex stainless steel with fine austenite spacing might become immune to hydrogen-induced cracking [15]. However, there is still empirical knowledge prevailing in modern engineering practices stating that duplex stainless steel components with austenite spacing smaller than 30 μm are recommended for extended service life in harsh applications [2, 16]. Furthermore, the root cause for the superior resistance to hydrogen-induced cracking of finely-grained microstructures has remained little understood. We believe the quasi-hydrides are precursor structures that precede or preclude hydrogen embrittlement, a question that has not been answered yet.

In this work, we shed more light on the diffraction peaks of the quasi-hydrides and corroborated the outcome with computational analyses. We performed ab-initio density functional theory (DFT) and thermodynamic calculations and explained the hydride formation mechanism. Finally, we revisit the view of hydrogen embrittlement and state that the degradation mechanism for stainless steel is phase decomposition due to phase instability caused by hydrogen absorption.

## Experimental and Computational Details

The material used was SAF 2507 SDSS with the following chemical composition (in wt.-%): 25.35% Cr, 6.46% Ni, 3.85% Mo, 0.44% Mn, 0.28% Si, 0.13% Cu, 0.29% N, and other trace elements. Miniature-sized tensile samples with a gauge length of 25 mm were fabricated. Our previous works provide a detailed description of the experimental setup [12, 13]. We report here only concise information.

*In-situ Grazing Incidence X-ray Diffraction*

The experiment setup has been explained in earlier communications [12, 13]. In short, GIXRD was employed to measure the diffraction patterns during electrochemical cathodic polarisation in situ. The diffraction measurements were carried out with a grazing angle of 0.05° to reveal lattice changes occurring in the near-surface region with a probing depth of approximately 40 nm, determined according to *Welzel et al.* [17]. The projected x-ray beam onto the surface was 2000 μm x 300 μm for the measurements with the grazing angle of 0.05°, producing diffraction signals from approximately 400-1200 grains. The projected beam area with a 0.2° grazing angle was 300 μm x 500 μm, producing diffraction signals from about 100-300 grains. The x-ray measurements were conducted at the beamline I07 at Diamond Light Source, using x-ray energy of 20.5 keV with a beam size of 100 μm (vertical) x 300 μm (horizontal) at the sample position. The experiments were performed with a DECTRIS Pilatus 100K two-dimensional detector with an area of 487 x 195 pixels (each 172 x 172 μm$^2$) mounted on the diffractometer arm at 900 mm from the sample. The tensile specimen was continuously polarised at a given electrochemical cathodic potential (all potentials vs Ag/AgCl (sat.)) while capturing XRD patterns from the surface. The potential was changed stepwise towards larger cathodic polarisation from OCP to -1.9 V without switching the potentiostat. The dwelling time for each step was approximately an hour or longer, so the electrochemical potentials were in a near-steady state. The diffraction measurements were done in two directions, in-plane and out-of-plane (see earlier works [12, 13]). As described previously, the 2D-diffraction data were converted into 1D patterns [18, 19].

*Thermodynamic Calculations using FactSage*

FactSage 8.1 software was used for calculating Pourbaix (E-pH) diagrams. Multicomponent models for the Fe-Cr-Ni-H$_2$O and Fe-Cr-Mo-H$_2$O systems were assessed using the E-pH module [20] at 298.15 K. $10^{-6}$ mol·kg$^{-1}$ was set as a standard value for all aqueous species for these systems. In the calculations, molar ratios of metallic elements were also fixed to predict the possible intermediate phases. We calculated Pourbaix diagrams with various Cr, Ni, and Mo concentrations. The compositions shifted the equilibria lines, and no other phase region formed. Therefore, we only report the Pourbaix diagram with 27% Cr and 7% Ni.

*Ab-initio Density Functional Theory Calculations*

We assume that the quasi-hydrides [13] are composed of iron and nickel, as FactSage calculations show Fe + $Ni_2H$ as the equilibrium conditions (shown later). Therefore, we have applied spin-polarized density functional theory (SP-DFT) to calculate the lattice parameters of $FeNi_3H_x$ quasi-hydrides and the defect formation energy of an interstitial hydrogen atom within the dilute solid solution limit in $FeNi_3$. The latter phase information was extracted from the Informatica Database (MedeA InfoMaticA version 3.1). The reason for investigating $FeNi_3$ is a nickel-rich surface alloy layer in stainless steel [21, 22]. The studied SDSS contains 7% nickel, and chemical and microstructural analyses show that the surface alloy layer could be composed of about 35-75% nickel with a thickness of 5-10 nm [21-23]. In most nickel-containing stainless steels, the surface oxide is composed of iron and chromium oxyhydroxide compounds due to selective oxidation, which enriches nickel beneath the passive film [24]. We have taken 75% Ni and 25% Fe for the composition of the surface alloy layer for clarity. It has been demonstrated that the diffraction patterns with lower Ni concentrations down to 20% are very similar [25]. Thus, $FeNi_3$ is a practical model for understanding eventual hydride evolution.

Considering the Fe-Ni binary diagram [26], there is $FeNi_3$ in all iron-nickel compositions. The diffraction patterns of $FeNi_3$ are practically the same as that of the austenite [25]. To investigate the effects of hydrogen charging on $FeNi_3$, we applied a generalized gradient approximation ~~used~~ as the exchange-correlation functional as parametrized by Perdew et al. (GGA-PBE) [27]. The projector augmented wave (PAW) method [28] was applied in the Vienna ab-initio simulation package (VASP code) [29] within the framework of the MedeA environment. The modelling was carried out in the reciprocal space, and plane-wave cut-off energy of 350 eV was applied, with a k-point spacing of 0.09 Å$^{-1}$. A First-order Methfessel-Paxton integration scheme [30] was used, with a smearing width of 0.1 eV. The self-consistency convergence criterion of 10$^{-4}$ eV was applied, and the atomic positions were relaxed until Hellman-Feynman forces were below 10 meV/Å. The formation energy of a defect ($\Delta H_f$) in a metallic lattice was computed according to Zhang and Northrup [31] as follows:

$$\Delta H_f = E_{tot}^q - \sum n_i \mu_i \quad \text{Eq. 1}$$

where $E_{tot}^q$ is the difference in the total formation energies of a defective and pristine $FeNi_3$ supercell ($E^{def} - E^{pure}$), $\mu_i$ is the chemical potential energy of the added hydrogen atoms to the supercell with $n_i$ being the number of added hydrogen atoms. Each added hydrogen atom's chemical potential was equilibrated to an environmental reservoir through the second term of Eq. 1. Thus, the defect formation energy of a single interstitial hydrogen atom ($\Delta H_f^{H_i}$) becomes:

$$\Delta H_f^{H_i} = E^{def} - E^{pure} - \mu_H \quad \text{Eq. 2}$$

Eq. 2 demonstrates the direct relationship between the formation energy of interstitial hydrogen and the chemical potential energy of hydrogen during electrochemical hydrogen charging. We have calculated $E^{def}$ under constant volume using periodic boundary conditions for a supercell size of 4×4×4, which corresponds to 256 atoms per supercell for a face-centred cubic (FCC) structure and 128 atoms per supercell for a body-centred cubic (BCC) structure, respectively. The hydrogen chemical potential, $\mu_H$ is influenced by many parameters, such as the electrolyte constituents, pH, cathodic potential, and any factor that impacts atomic hydrogen concentration and thermodynamic activity at the surface [32]. The definition of $\mu_H$ is as follows:

$$\mu_H = \partial G / \partial n_H \qquad \text{Eq. 3}$$

where $G$ is the Gibbs free energy and $n_H$ is the number of hydrogen atoms within the phase at equilibrium. $\mu_H$ can be directly linked to thermodynamic activity at a given temperature T:

$$\mu_H = E_{ref} + \mu^0(T) + kT\ln(a) \qquad \text{Eq. 4}$$

$E_{ref}$ is reference hydrogen reservoir energy, which depends on experimental conditions, $k$ is the Boltzmann constant, $T$ is the absolute temperature, and $a$ is the hydrogen activity, which was considered 1. Hence, during hydrogen charging, an increase in $\mu_H$ is inevitable due to a substantial increase in hydrogen activity [33, 34].

**Results and Discussion**

**Figure 1** summarises the XRD measurement results for the austenite (111) and ferrite (110) Bragg peaks. The austenite peaks split in two when the hydrogen reduction potential (about -850 mV vs standard hydrogen electrode, i.e., -1070 mV vs Ag/AgCl) was exceeded, upon which excessive hydrogen absorption took place. The extent of the peak splitting increased as a function of the magnitude of cathodic polarisation potential, indicating more hydrogen uptake. As demonstrated in an earlier communication, peak splitting occurred for all austenite peaks associated with hydrogen charging [13]. We have shown that hydrogen in the austenite and ferrite lattice causes strain, which increases with the uptaken hydrogen [9, 11-13]. The parent austenite peak almost disappeared at -1.9 V, and the new Bragg signal became dominating. This phenomenon is shown in more clarity in **Figure 2**. There was excessive hydrogen evolution beyond -1.7 V since a continuous large stream of hydrogen bubble formation became visible to the naked eye. Both half detector images show discrete Bragg peaks with heterogeneous strain development associated with hydrogen charging. Hence, hydrogen-induced strain evolution occurs more readily on specific or large grains.

**Figure 3** shows the out-of-plane diffraction data obtained at increasing cathodic polarisation and calculated (DFT) diffraction data of nickel hydride ($Ni_2H$). The signal intensities of the austenite peaks increased with increasing cathodic polarisation while the peaks of the ferrite phase reduced. The volume interaction of the incoming x-rays was constant, so a change in intensity is a clear sign of the phase ratio change. Moreover, polarisation at -1.5 V produced additional peaks at around 15 and 23 degrees 2 Theta. These peaks were broad and indicated two or more and/or defective and nanocrystalline structures. DFT and FactSage calculations predict the formation of nickel hydrides (to be shown later). DFT calculated diffraction peaks of $Ni_2H$ showed a good match with the measured diffraction peaks.

The DFT calculations have shown that the formation energy of an interstitial hydrogen defect in the (nickel-rich) austenite phase is significantly lower than in the (nickel-rich) ferrite phase, irrespective of hydrogen chemical potential (**Figure 4**). The calculations further show that the interaction of hydrogen with the austenite is more intense than the ferrite, as the concentration of hydrogen interstitials depends exponentially on defect formation energy based on an Arrhenius relation (Eq. 5) [34]. Hydrogen interstitials cause an increase in lattice parameters and thus induce significant stress. Therefore, the immense strain development in the austenite is justified by its greater $H_i$ concentrations (due to lower $\Delta H_f^{H_i}$). The analyses show that the solubility of hydrogen increases exponentially with an increase in hydrogen activity ($\mu_H$), indicating the possibility of hydrogen occupation at all possible interstitial sites, given the rise in $\mu_H$. The latter may lead to the decomposition of the austenite phase. With an increase in the thermodynamic activity of hydrogen (or rise in the chemical potential of hydrogen; see Equation 5) beyond -3.7 eV, the defect formation energy of a hydrogen interstitial in the lattice becomes negative. A negative defect formation energy implies that the defect concentration exceeds the number of available atomic sites, making the phase unstable [35]. A negative defect formation energy also means spontaneous emergence of the dopant (interstitial hydrogen defect) beyond the anticipated solubility limit (208 ppm of hydrogen in the austenite [36]) even under non-equilibrium conditions, as demonstrated by Sun et al. [37].

The latter is better understood in Eq. 5, where $[d]_{H_i}$ is the defect concentration, $N$ represents the total number of possible sites, $T$ is the absolute temperature, and $k$ is the Boltzman constant. If $\Delta H_f^{H_i}$ becomes negative, the value of $[d]_{H_i}$ exceeds N, which reasons phase decomposition.

$$[d]_{H_i} = N \, exp\left(\frac{-\Delta H_f^{H_i}}{kT}\right) \qquad \text{Eq.5}$$

However, kinetic constraints can be decisive. The austenite phase is not stable under (excessive) hydrogen charging, and an ultrahigh quantity of hydrogen interstitials is formed throughout the hydrogen absorption process. Thus, the austenite phase eventually decomposes into a new quasi-phase, primarily an iron-nickel hydride type. Our statements align with the report of *Hinotani et al.* [38], who observed nickel hydrides during electrochemical hydrogen charging. In super duplex stainless steel, nickel has the strongest affinity to hydrogen among all other elements. Therefore, the potential hydride must contain nickel. However, hydride evolution can't exclude chromium and other alloying elements. Our aim is not to provide exact compositions but to shed light on the quasi-hydride formation mechanism. FactSage predicts a variety of hydrides, including chromium and iron hydrides, for the alloying system, but only nickel hydrides as the most stable ones. We have added detailed FactSage calculation results to the supplementary for further investigation.

The DFT calculations have shown the possible emergence of various $FeNi_3H_x$ quasi-hydrides, where $x$ depends on local hydrogen activity (suggesting no fixed stoichiometry). The latter explains the observed diffraction patterns with their variations in Bragg peak positions. However, FactSage calculations show that under equilibrium conditions, the stable phases are primarily Fe and $Ni_2H$ (**Figure 5**), which is further justified by our DFT analyses that demonstrate the phase decomposition of nickel-rich austenite under conditions of $\mu_H > -3.7$ eV (**Figure 5**). It is noteworthy that this range of $\mu_H$ corresponds to phase decomposition of austenite even under $\mu_H^{H_2}$ (the hydrogen chemical potential referenced to hydrogen molecule; $\mu_H = -3.39$ eV based on the GGA-PBE functional). The latter also explains why phase decomposition of the ferrite phase is less favoured than the austenite phase.

The hydride formation in the ferrite is more energy-consuming and requires a higher $\mu_H$; therefore, it is not as favoured. The required increase in hydrogen activity for the decomposition of the ferrite phase is significantly higher than the austenitic phase. Thus, the defect formation energy becomes negative only if the hydrogen chemical potential exceeds -3.2 eV (**Figure 4**), which is less likely to be reached. This argument is consistent with our experimental observations showing that the extensive XRD patterns change mainly in the austenite and not in the ferrite. However, the calculations show a similar fate awaiting the ferrite if the hydrogen chemical potential increases to extreme values. The results also clarify why the metastable phase formation (XRD peak splitting) is reversible: when hydrogen absorption terminates, the thermodynamic activity of hydrogen reduces, thereby reducing the hydrogen chemical potential, which consequently increases the defect formation energy of hydrogen interstitials; thus, the solubility of hydrogen decreases exponentially leading to an outflow (or desorption) of hydrogen from the lattice. Hence, the austenite phase becomes re-stabilized.

The calculations and experiments show that the hydrogen absorption process is reversible to a large extent. However, we have found plastic deformation upon electrochemical hydrogen charging of the studied alloy, indicating that hydrogen promoted dislocation activities and hence slip plane motions during the absorption of hydrogen into the microstructure [9]. Most of the lattice strains are elastic and therefore reversible, but hydrogen absorption resulted in irreversible lattice change on the order of $10^{-3}$. Moreover, hydrogen softens the austenite while the ferrite is hardened [10]. The data suggests that the embrittlement manifestation of the austenite phase is due to hydrogen-enhanced localized plasticity (HELP), in line with previous works [9-14]. In contrast, hydrogen-enhanced decohesion (HEDE) operates in the ferrite phase. Embrittlement due to hydride formation has been known for stable hydrides in titanium and nickel alloys [39]. However, hydride-induced cracking has, to our knowledge, at least, not been reported earlier in stainless steel. The observed hydrides are metastable and reversible; hence, their effect on possible embrittlement remains to be elaborated. The austenite phase forms the quasi-hydrides, but the ferrite phase typically suffers from hydrogen-induced brittle cracking [5, 40]. Therefore, the presence of the hydrides might delay cracking, which could be understood as a positive outcome (improved resistance to hydrogen embrittlement).

Hydrogen absorption results in an increase in the lattice parameters (**Figure 1**). The computed lattice parameters of various $FeNi_3H_x$ quasi-hydrides and their respective changes in diffraction peaks to the austenite phase are listed in **Table 1**. Such structures will inevitably result in strain due to increased unit cell volume of austenite upon hydrogen ingress. We calculated the diffraction patterns of various compositions containing nickel, iron and chromium, and none of the results gave a good match like the ones shown here. Therefore, we provided a concise summary of the calculations.

We have calculated the studied steel's electrochemical potential – pH diagram with different nickel equivalents to provide insight into possible physicochemical processes occurring on the surface (**Figure 5** and **Table 2**). We have calculated potential – pH diagrams with various nickel concentrations but only reported 27% Cr and 7% Ni. It has been reported that metallic nickel extends the stability regions of chromium oxide and iron-chromium oxide, improving the alloy's ability to passivation. The alloy's composition is 7 wt.-% Ni. However, nickel is enriched in the austenite phase by about 8 wt.-% and exists at lesser compositions in the ferrite phase at about 5-6 wt.-%. The surface comprises an outer layer of iron oxide/hydroxide and an inner chromium oxide/hydroxide, which are about 2 nm thick [21, 22, 41].

Furthermore, there is a nickel-rich alloy surface layer below the surface oxides, whose composition has not been clarified yet [21, 22, 41]. In light of our previous works [21, 22, 41] and the results of others [42, 43], we assume the nickel composition of the surface alloy layer of about 50-75 wt.-%. There is more nickel in austenitic grains than ferritic ones, and the total thickness is nearly twofold the thickness of the oxide layer. We have shown in earlier work that the surface alloy layer exists in 25Cr-7Ni super duplex stainless steel and that its composition can change depending on the electrode's potential [22]. Hence, we speculate similar behaviour in this work and anticipate compositional and structural (thickness) changes of the oxides during electrochemical polarisation. The NaCl solution (0.1 M) used for electrochemical hydrogen charging was near-neutral. We used Ag/AgCl (saturated with KCl) as a reference electrode. Therefore, the potentials in **Figure 5** are shifted by approximately 220 mV to lower values. The OCP was about 0 mV. The potential – pH diagram shows the presence of nickel, iron, and chromium compounds at pH 7 from OCP (-0.2 V) to cathodic potentials down to -1.1 V. Corrosion is impeded by the presence of the passive surface film, i.e. chromium oxide/hydroxide. So, the peak shifts cannot be explained by corrosion grounds. Corrosion is an anodic reaction (Faradaic) leading to metal dissolution and/or formation of corrosion products on the surface. During the cathodic polarisation at large negative potentials, metal dissolution is not likely, so the peak shifts can't be explained by corrosion processes. Nevertheless, we measured the metal concentration in the electrolyte after hydrogen charging by atomic emission spectroscopy and no element of the alloy's composition was detected, corroborating the earlier statement.

The potential-pH diagram shows stable nickel hydride ($Ni_2H$) under cathodic potentials at all pH. DFT calculations also show the feasible formation of metastable quasi-nickel hydrides, and their calculated diffraction patterns matched well with the measured ones. The diffraction experiments showed that the second peak at lower 2-Theta (hydride, see **Figure 2**) changed in position, full width at half maximum and intensity with charging time and potential, indicating that the new structure is relatively active and metastable. Factstage calculations show the thermodynamically stable hydride is only $Ni_2H$. We polarised for a total time of a few hours. The phase transformation has a kinetic constraint since the peak continuously changes with cathodic potential and charging time, suggesting that the phase transformation from austenite to hydride depends on the hydrogen and nickel concentration and their respective chemical potentials. The more the two, the faster hydride evolution occurs, in line with the report of Hinotani et al. [38]. The surface always contains the most hydrogen and has a nickel-rich surface alloy layer. So, we believe the phase transformation begins at the surface beneath the oxide, which still exists throughout the hydrogen charging process. When the hydrogen concentration reaches the point of saturation, the lattice becomes unstable. Hence, the lattice can dissolve more hydrogen and transform into a metastable hydride. Factstage calculation suggests that the nickel-rich hydride becomes a stable pure nickel hydride after some time. The phase transformation process is schematically illustrated in **Figure 6**. However, the austenite regains stability when the hydrogen charging process is terminated, reasoning a decrease in hydrogen activity.

It should be emphasized that FactSage calculation predicts pure nickel hydride, metallic iron, and metallic chromium at cathodic potentials far below the hydrogen evolution potential (< -1.1 V). This outcome strongly suggests that nickel is the hydride former, although iron and chromium can also form hydrides. Therefore, we do not exclude iron, chromium and other alloying elements participating in the hydride lattice in the austenite phase, possibly creating a mixed-typed quasi-hydride. Still, the thermodynamic stable hydride is solely $Ni_2H$, regardless of nickel concentration. Furthermore, iron is metallic, indicating that all iron oxyhydroxides reduce after long-term cathodic polarisation. Chromium oxide is more resistant to cathodic polarisation than iron oxides.

Since the pH near the surface becomes alkaline during hydrogen charging, it can provide a barrier for hydrogen absorption, thereby delaying hydrogen embrittlement. However, chromium oxide is reduced to metallic chromium at low potentials. Interestingly, FactSage predicts no hydrogen in Fe and Cr once $Ni_2H$ is formed.

## Conclusions

Our results have shown that continuous hydrogen absorption destabilizes the microstructure. Eventual hydride formation becomes possible when there is enough hydrogen absorbed. The reduction of diffraction peak intensities of the ferrite phase and the intensity rise of the austenite peaks suggest that there is also phase decomposition of the ferrite phase to FCC crystal structure, i.e., nickel hydrides. DFT and FactSage calculations show possible thermodynamic structures without providing kinetic information. However, the GIXRD measurements suggest that the phase decomposition of the ferrite phase is delayed. Hydrogen degrades the microstructure, causing embrittlement at its final stage. Our findings report the most nascent stages of microstructural degradation. Hence, hydrogen is a phase decomposer. It operates more efficiently in the austenite phase than in the ferrite phase. We emphasize that the phase decomposition is reversible, as suggested by FactSage entirely. Upon termination of hydrogen charging (cathodic polarisation), the metastable phases revert to the initial stages.

Hence, the post-mortem characterization of an eventually fractured piece of stainless steel does not honestly picture what happened. Hydrogen embrittlement becomes only understood if measured operando and in real-time, as shown above. We revisit the theory of hydrogen embrittlement and state that hydrogen degrades the microstructure by destabilizing it, causing phase decomposition into hydrides and their pure metallic structures. The classic understanding that hydrogen causes lattice decohesion or localized plasticity is undoubtedly correct but is more fracture mechanical aspect of degradation. The reason for hydrogen embrittlement is instead associated with a compositionally and structurally metastable alloy. The formation of hydrogen-induced cracks is anticipated on sites where the phase decomposition occurs locally in the ferrite and austenite phases. In stress, phase decomposition can be accelerated; hydrogen embrittlement appears in areas with the highest stress and most propensity for phase separation. Our hypothesis explains microstructure softening due to hydrogen absorption [9, 10]. The pure phases that form during hydrogen charging have lower elastic moduli than the alloy phases; hence, dislocation mobility is enhanced, reasoning softening.

## Data availability

All data needed to evaluate the conclusions in the paper are present in the article, and further data is provided in the Supplementary Materials. Additional data related to this paper or the raw data may be obtained from the authors upon a reasonable request.

## CRediT authorship contribution statement

**Cem Örnek**: Methodology, Software, Validation, Formal analysis, Investigation, Resources, Data curation, Writing – original draft, Writing – review & editing, Visualization, Supervision, Project administration, Funding acquisition. **Alfred Larsson**: Software, Formal analysis, Investigation,

Resources, Data curation, Writing – original draft, Writing – review & editing. **Mubashir Mansoor**: Software, Formal analysis, Investigation, Resources, Data curation, Writing – original draft, Writing – review & editing, Visualization. **Fan Zhang**: Investigation, Writing – review & editing. **Gary S Harlow**: Investigation, Writing – review & editing. **Robin Kroll**: Investigation, Resources, Writing – review & editing. **Francesco Carlà**: Conceptualization, Methodology, Investigation, Resources, Writing – review & editing, Supervision. **Hadeel Hussain**: Investigation, Resources, Writing – review & editing. **Bora C Derin**: Software, Formal analysis, Investigation, Resources, Data curation, Writing – review & editing. **Ulf Kivisäkk**: Resources, Writing – review & editing. **Dirk L Engelberg**: Investigation, Resources, Writing – review & editing, Supervision, Funding acquisition. **Edvin Lundgren**: Resources, Writing – review & editing, Supervision, Funding acquisition. **Jinshan Pan**: Conceptualization, Methodology, Validation, Investigation, Resources, Writing – review & editing, Supervision, Project administration, Funding acquisition.

## Competing interests

The authors declare that they have no competing interests.

## Acknowledgements

This work was supported by TÜBITAK (The Scientific and Technological Research Council of Turkey) under contract number 118C227 within the program *2232: International Fellowship for Outstanding Researchers*, by Vinnova (Sweden's innovation agency) under contract number 2018–03267 within the Vinnova program of *Industrial pilot projects for neutron and photon experiments at large scale research infrastructures*, and by the Swedish Research Council within the program Röntgen-Ångström Cluster *In-situ High Energy X-ray Diffraction from Electrochemical Interfaces (HEXCHEM)* under the project no. 2015–06092. This work was also performed with partial support from the Nanometer Structure Consortium at Lund University (nmC@LU). Furthermore, we acknowledge Diamond Light Source for time on beamline I07 under proposal SI23388. Finally, we thank Zuhal Er (ITU) for the computational support at the National Center for High-Performance Computing of the Republic of Turkey (UHeM) under Grant No. 1008852020. Cem Örnek and Mubashir Mansoor are grateful to Mustafa Ürgen, Istanbul Technical University, for fruitful discussions. Mubashir Mansoor appreciates conversations with Kamil Czelej at the University of California about Density Functional Theory calculations.

## Appendix A. Supplementary data

The following are the supplementary data to this article:

## References

[1] H.K.D.H. Bhadeshia, Prevention of Hydrogen Embrittlement in Steels, ISIJ International, 56 (2016) 24-36.
[2] G. Byrne, R. Francis, G. Warburton, Hydrogen Induced Stress Cracking (HISC) Resistance and Improvement Methods for Super Duplex Stainless Steels - Paper No. 6981, in: Corrosion 2016, NACE International, Vancouver, British Columbia, Canada, 2016.

[3] M.S. Hazarabedian, A. Viereckl, Z. Quadir, G. Leadbeater, V. Golovanevskiy, S. Erdal, P. Georgeson, M. Iannuzzi, Hydrogen-induced stress cracking of swaged super duplex stainless steel subsea components, corrosion, 75 (2019) 824-838.

[4] O. Barrera, D. Bombac, Y. Chen, T.D. Daff, E. Galindo-Nava, P. Gong, D. Haley, R. Horton, I. Katzarov, J.R. Kermode, C. Liverani, M. Stopher, F. Sweeney, Understanding and mitigating hydrogen embrittlement of steels: a review of experimental, modelling and design progress from atomistic to continuum, Journal of Materials Science, 53 (2018) 6251-6290.

[5] R. Oltra, C. Bouillot, T. Magnin, Localized hydrogen cracking in the austenitic phase of a duplex stainless steel, Scripta Materialia, 35 (1996) 1101-1105.

[6] M. Dadfarnia, M.L. Martin, A. Nagao, P. Sofronis, I.M. Robertson, Modeling hydrogen transport by dislocations, Journal of the Mechanics and Physics of Solids, 78 (2015) 511-525.

[7] S. Wang, M.L. Martin, P. Sofronis, S. Ohnuki, N. Hashimoto, I.M. Robertson, Hydrogen-induced intergranular failure of iron, Acta Materialia, 69 (2014) 275-282.

[8] I.M. Robertson, P. Sofronis, A. Nagao, M.L. Martin, S. Wang, D.W. Gross, K.E. Nygren, Hydrogen Embrittlement Understood, Metallurgical and Materials Transactions B, 46 (2015) 1085-1103.

[9] C. Örnek, T. Müller, B.M. Şeşen, U. Kivisäkk, F. Zhang, M. Långberg, U. Lienert, A. Jeromin, T.F. Keller, E. Lundgren, J. Pan, Hydrogen-Induced Micro-Strain Evolution in Super Duplex Stainless Steel—Correlative High-Energy X-Ray Diffraction, Electron Backscattered Diffraction, and Digital Image Correlation, Frontiers in Materials, 8 (2022).

[10] C. Örnek, B.M. Şeşen, M.K. Ürgen, Understanding Hydrogen-Induced Strain Localization in Super Duplex Stainless Steel Using Digital Image Correlation Technique, Met. Mater. Int., (2021).

[11] C. Örnek, T. Müller, U. Kivisäkk, F. Zhang, M. Långberg, U. Lienert, K.-H. Hwang, E. Lundgren, J. Pan, Operando time- and space-resolved high-energy X-ray diffraction measurement to understand hydrogen-microstructure interactions in duplex stainless steel, Corrosion Science, 175 (2020) 108899.

[12] C. Örnek, A. Larsson, G.S. Harlow, F. Zhang, R. Kroll, F. Carlà, H. Hussain, U. Kivisäkk, D.L. Engelberg, E. Lundgren, J. Pan, Time-resolved grazing-incidence X-ray diffraction measurement to understand the effect of hydrogen on surface strain development in super duplex stainless steel, Scripta Materialia, 187 (2020) 63-67.

[13] C. Örnek, A. Larsson, G.S. Harlow, F. Zhang, R. Kroll, F. Carlà, H. Hussain, U. Kivisäkk, D.L. Engelberg, E. Lundgren, J. Pan, Metastable Precursor Structures in Hydrogen-infused Super Duplex Stainless Steel Microstructure – An Operando Diffraction Experiment, Corrosion Science, (2020) 109021.

[14] C. Örnek, P. Reccagni, U. Kivisäkk, E. Bettini, D.L. Engelberg, J. Pan, Hydrogen embrittlement of super duplex stainless steel – Towards understanding the effects of microstructure and strain, International Journal of Hydrogen Energy, 43 (2018) 12543-12555.

[15] F. Vucko, G. Ringot, D. Thierry, N. Larché, Fatigue Behavior of Super Duplex Stainless Steel Exposed in Natural Seawater Under Cathodic Protection, Frontiers in Materials, 9 (2022).

[16] D.N.V. (DNV), Recommended Practice DNV-RP-F112 Design of Duplex Stainless Steel Subsea Equipment Exposed to Cathodic Protection, in, Det Norkse Veritas, Norway, 2008.

[17] U. Welzel, J. Ligot, P. Lamparter, A.C. Vermeulen, E.J. Mittemeijer, Stress analysis of polycrystalline thin films and surface regions by X-ray diffraction, Journal of Applied Crystallography, 38 (2005) 1-29.

[18] C.M. Schlepütz, S.O. Mariager, S.A. Pauli, R. Feidenhans'l, P.R. Willmott, Angle calculations for a (2+3)-type diffractometer: focus on area detectors, Journal of Applied Crystallography, 44 (2011) 73-83.

[19] G. Abbondanza, A. Larsson, F. Carla, E. Lundgren, G.S. Harlow, Quantitative powder diffraction using a (2 + 3) surface diffractometer and an area detector, Journal of Applied Crystallography, 54 (2021) 1140-1152.

[20] C.W. Bale, E. Bélisle, P. Chartrand, S.A. Decterov, G. Eriksson, A.E. Gheribi, K. Hack, I.H. Jung, Y.B. Kang, J. Melançon, A.D. Pelton, S. Petersen, C. Robelin, J. Sangster, P. Spencer, M.A. Van Ende, FactSage thermochemical software and databases, 2010–2016, Calphad, 54 (2016) 35-53.

[21] M. Långberg, F. Zhang, E. Grånäs, C. Örnek, J. Cheng, M. Liu, C. Wiemann, A. Gloskovskii, T.F. Keller, C. Schlueter, S. Kulkarni, H. Noei, D. Lindell, U. Kivisäkk, E. Lundgren, A. Stierle, J. Pan, Lateral variation of the native passive film on super duplex stainless steel resolved by synchrotron hard X-ray photoelectron emission microscopy, Corrosion Science, 174 (2020) 108841.

[22] M. Långberg, C. Örnek, J. Evertsson, G.S. Harlow, W. Linpé, L. Rullik, F. Carlà, R. Felici, E. Bettini, U. Kivisäkk, E. Lundgren, J. Pan, Redefining passivity breakdown of super duplex stainless steel by electrochemical operando synchrotron near surface X-ray analyses, npj Materials Degradation, 3 (2019) 22.

[23] R.P. Oleksak, R. Addou, B. Gwalani, J.P. Baltrus, T. Liu, J.T. Diulus, A. Devaraj, G.S. Herman, Ö.N. Doğan, Molecular-scale investigation of the oxidation behavior of chromia-forming alloys in high-temperature $CO_2$, npj Materials Degradation, 5 (2021) 46.

[24] C.O.A. Olsson, D. Landolt, Passive films on stainless steels—chemistry, structure and growth, Electrochimica Acta, 48 (2003) 1093-1104.

[25] H. Lu, Z. Liu, X. Yan, D. Li, L. Parent, H. Tian, Electron work function–a promising guiding parameter for material design, Scientific Reports, 6 (2016) 24366.

[26] C.W. Yang, D.B. Williams, J.I. Goldstein, A revision of the Fe-Ni phase diagram at low temperatures (<400 °C), Journal of Phase Equilibria, 17 (1996) 522-531.

[27] J.P. Perdew, K. Burke, M. Ernzerhof, Generalized Gradient Approximation Made Simple, Physical Review Letters, 77 (1996) 3865-3868.

[28] P.E. Blöchl, Projector augmented-wave method, Physical Review B, 50 (1994) 17953-17979.

[29] G. Kresse, D. Joubert, From ultrasoft pseudopotentials to the projector augmented-wave method, Physical Review B, 59 (1999) 1758-1775.

[30] M. Methfessel, A.T. Paxton, High-precision sampling for Brillouin-zone integration in metals, Physical Review B, 40 (1989) 3616-3621.

[31] S.B. Zhang, J.E. Northrup, Chemical potential dependence of defect formation energies in GaAs: Application to Ga self-diffusion, Physical Review Letters, 67 (1991) 2339-2342.

[32] R. Nazarov, T. Hickel, J. Neugebauer, First-principles study of the thermodynamics of hydrogen-vacancy interaction in fcc iron, Physical Review B, 82 (2010) 224104.

[33] S.W. Boettcher, S.Z. Oener, M.C. Lonergan, Y. Surendranath, S. Ardo, C. Brozek, P.A. Kempler, Potentially Confusing: Potentials in Electrochemistry, ACS Energy Letters, 6 (2021) 261-266.

[34] C. Freysoldt, B. Grabowski, T. Hickel, J. Neugebauer, G. Kresse, A. Janotti, C.G. Van de Walle, First-principles calculations for point defects in solids, Reviews of Modern Physics, 86 (2014) 253-305.

[35] N. Wang, D. West, J. Liu, J. Li, Q. Yan, B.-L. Gu, S.B. Zhang, W. Duan, Microscopic origin of the $p$-type conductivity of the topological crystalline insulator SnTe and the effect of Pb alloying, Physical Review B, 89 (2014) 045142.

[36] A. Turk, S.D. Pu, D. Bombač, P.E.J. Rivera-Díaz-del-Castillo, E.I. Galindo-Nava, Quantification of hydrogen trapping in multiphase steels: Part II – Effect of austenite morphology, Acta Materialia, 197 (2020) 253-268.

[37] W. Sun, A. Holder, B. Orvañanos, E. Arca, A. Zakutayev, S. Lany, G. Ceder, Thermodynamic Routes to Novel Metastable Nitrogen-Rich Nitrides, Chemistry of Materials, 29 (2017) 6936-6946.

[38] S. Hinotani, Y. Ohmori, F. Terasaki, Effect of nickel on hydride formation and hydrogen embrittlement in Ni-Cr-Fe alloys, Materials Science and Engineering, 74 (1985) 119-131.

[39] S. Lynch, Hydrogen embrittlement phenomena and mechanisms, Corrosion Reviews, 30 (2012) 105-123.

[40] B. An, T. Iijima, C. San Marchi, B. Somerday, Micromechanisms of Hydrogen-Assisted Cracking in Super Duplex Stainless Steel Investigated by Scanning Probe Microscopy, in: ASME 2014 Pressure Vessels and Piping Conference, 2014.

[41] M. Långberg, C. Örnek, F. Zhang, J. Cheng, M. Liu, E. Grånäs, C. Wiemann, A. Gloskovskii, Y. Matveyev, S. Kulkarni, H. Noei, T.F. Keller, D. Lindell, U. Kivisäkk, E. Lundgren, A. Stierle, J. Pan, Characterization of Native Oxide and Passive Film on Austenite/Ferrite Phases of Duplex Stainless Steel Using Synchrotron HAXPEEM, Journal of The Electrochemical Society, 166 (2019) C3336-C3340.

[42] X.X. Wei, B. Zhang, B. Wu, Y.J. Wang, X.H. Tian, L.X. Yang, E.E. Oguzie, X.L. Ma, Enhanced corrosion resistance by engineering crystallography on metals, Nature Communications, 13 (2022) 726.

[43] P. Marcus, V. Maurice, The Structure of Passive Films on Metals and Alloys, in: M.B. Ives, J.L. Luo, J.R. Rodda (Eds.) Passivity of Metals and Semiconductors, The Electrochemical Society, Jasper Park Lodge, Canada, 1999, pp. 30-64.

# Tables

**Table 1:** *The variations in lattice parameters, density, and change in Bragg's angle (energy = 20.5 keV) for various quasi-hydrides relative to the nickel-rich austenite phase are shown.*

| Structure | Lattice parameters (Å) | Density (gm/cm$^3$) | Change in Bragg angle (2θ) |
|---|---|---|---|
| FeNi$_3$ | 3.542 | 8.669 | 0 |
| FeNi$_3$H$_{0.015}$ | 3.542 | 8.669 | - 0.001 |
| FeNi$_3$H$_{0.125}$ | 3.553 | 8.590 | - 0.060 |
| FeNi$_3$H | 3.606 | 8.248 | - 0.310 |

**Table 2:** *Stable species for the calculated potential-pH diagram.*

| No. | Species | DG (kJ) | DG (kcal) |
|---|---|---|---|
| 35 | Cr[2+] | -176.183 | -42.109 |
| 36 | Cr[3+] | -215.748 | -51.565 |
| 37 | CrO$_4$[2-] | -727.907 | -173.974 |
| 38 | Cr$_2$O$_7$[2-] | -1301.501 | -311.066 |
| 39 | Cr(OH)[2+] | -430.866 | -102.979 |
| 40 | HCrO$_4$[-] | -764.897 | -182.815 |
| 41 | Fe[2+] | -78.874 | -18.851 |
| 42 | Fe[3+] | -4.612 | -1.102 |
| 43 | FeOH[+] | -277.247 | -66.264 |
| 44 | FeOH[2+] | -229.558 | -54.866 |
| 45 | Fe$_2$(OH)$_2$[4+] | -467.197 | -111.663 |
| 46 | Ni[2+] | -45.598 | -10.898 |
| 47 | NiOH[+] | -227.621 | -54.403 |
| 48 | HNiO$_2$[-] | -344.207 | -82.268 |
| 49 | Cr(s) | 0 | 0 |
| 50 | CrO$_2$(s) | -528.644 | -126.349 |
| 51 | CrO$_3$(s) | -500.968 | -119.734 |
| 52 | Cr$_2$O$_3$(s) | -1046.658 | -250.157 |
| 53 | Cr(CrO$_4$)(s) | -1058.784 | -253.056 |
| 54 | Cr$_3$O$_4$(s) | -1324.656 | -316.6 |
| 55 | Cr$_2$(CrO$_4$)$_3$(s) | -2660.18 | -635.798 |
| 56 | Cr$_8$O$_{21}$(s) | -4132.61 | -987.718 |
| 57 | Cr(OH)$_2$(s) | -507.694 | -121.342 |
| 58 | Fe(s) | 0 | 0 |
| 59 | Fe(s2) | 5.402 | 1.291 |
| 60 | FeO(s) | -244.871 | -58.526 |
| 61 | Fe$_2$O$_3$(s) | -743.978 | -177.815 |
| 62 | Fe$_2$O$_3$(s2) | -668.149 | -159.692 |
| 63 | Fe$_2$O$_3$(s3) | -663.114 | -158.488 |
| 64 | Fe$_3$O$_4$(s) | -1014.302 | -242.424 |
| 65 | Fe$_3$O$_4$(s2) | -1005.622 | -240.349 |
| 66 | Fe$_3$O$_4$(s3) | -853.509 | -203.994 |
| 67 | Fe$_3$O$_4$(s4) | -844.827 | -201.919 |
| 68 | Fe(OH)$_2$(s) | -492.047 | -117.602 |
| 69 | Fe(OH)$_3$(s) | -705.577 | -168.637 |
| 70 | Fe$_2$O$_3$(H2O)(s) | -975.966 | -233.261 |
| 71 | FeCr$_2$O$_4$(s) | -1356.014 | -324.095 |
| 72 | Ni(s) | 0 | 0 |
| 73 | Ni$_2$H(s) | 11.492 | 2.747 |

| 74 | NiO(s) | -211.748 | -50.609 |
| 75 | NiOOH(s) | -319.778 | -76.429 |
| 76 | $Ni(OH)_2$(s) | -446.946 | -106.823 |
| 77 | $NiO_2(H_2O)$(s) | -409.922 | -97.974 |
| 78 | $(NiO)(Cr_2O_3)$(s) | -1267.59 | -302.961 |
| 79 | $(NiO)(Fe_2O_3)$(s) | -972.982 | -232.548 |
| 80 | $(NiO)(Fe_2O_3)$(s2) | -972.42 | -232.414 |

# Figures

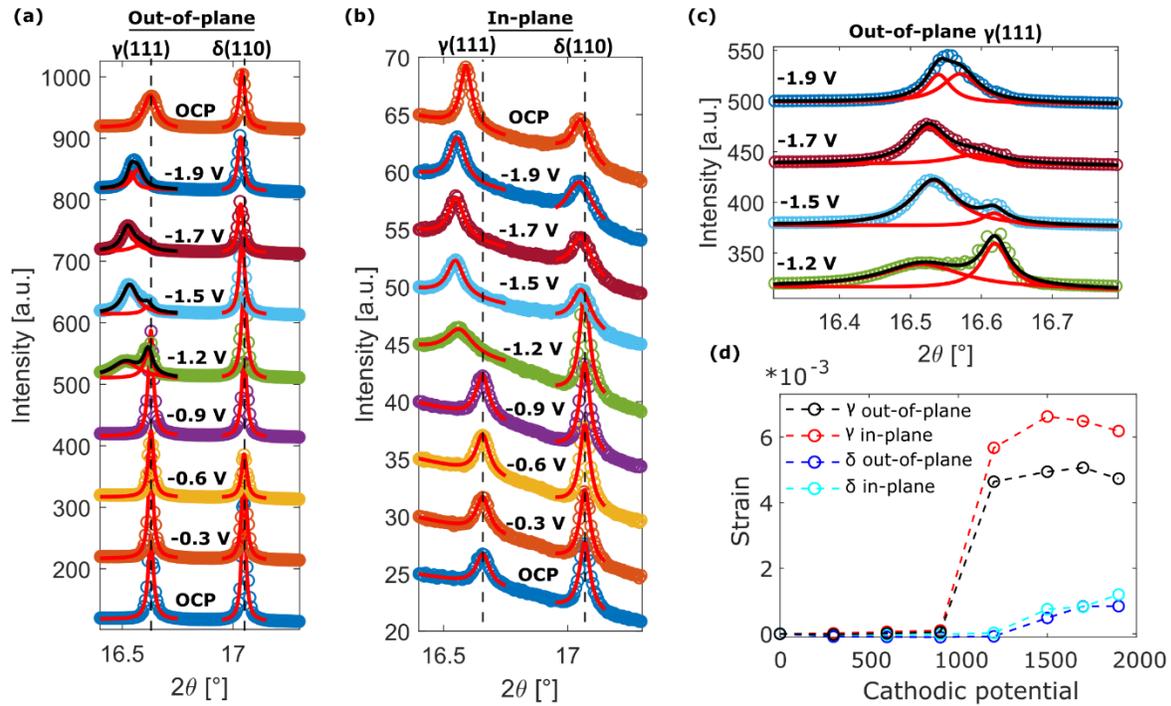

*Figure 1:* *Operando GIXRD results showing the collected diffraction patterns of the ferrite phase and the austenite phase: (a) out-of-plane (along the specular direction) and (b) in-plane (90° off the specular direction) vectors, (c) magnified view of the austenite phase for high cathodic potentials, and (d) showing the strain evolution. The out-of-plane diffraction signals emanate nearly perpendicular to the sample surface, while the in-plane direction is parallel to the surface. Note that an Ag/AgCl reference electrode was used, so the potential values are 0.22 V more negative than the standard hydrogen electrode.*

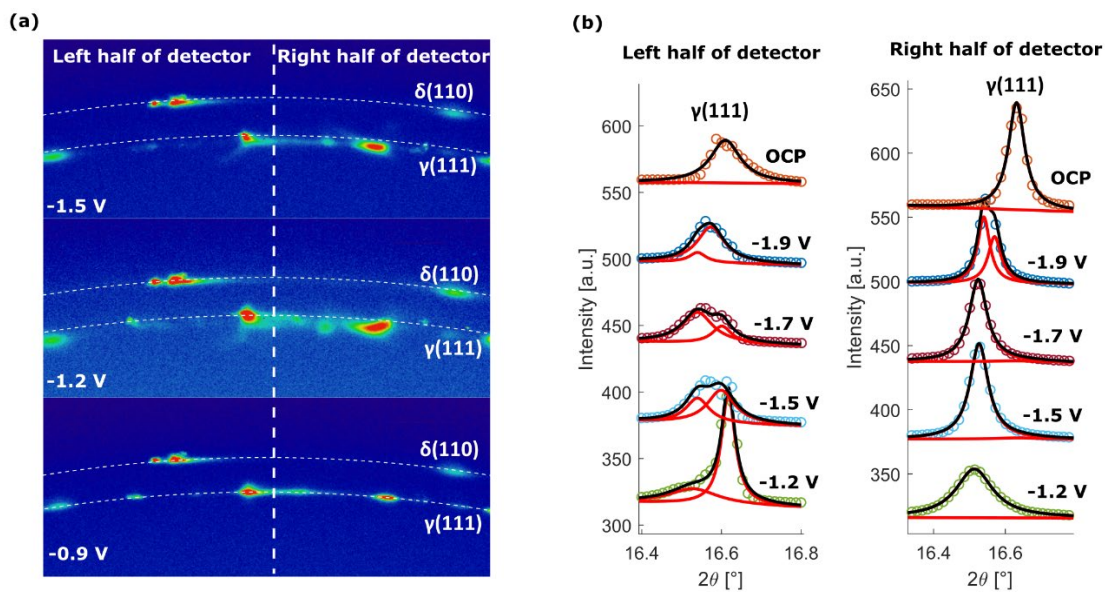

*Figure 2:* Out-of-plane diffraction data obtained at increasing cathodic potentials displaying paternal changes of the austenite and ferrite phases, (b) integrated 1D XRD pattern for the left and right detector halves plotted against the cathodic potentials.

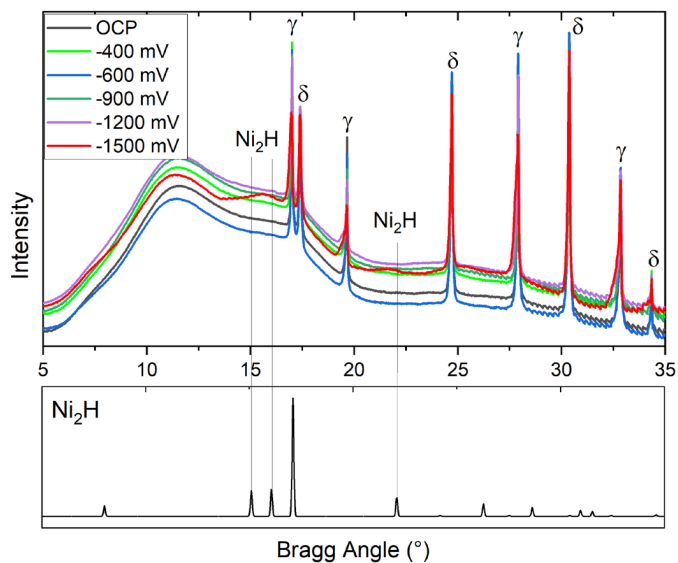

*Figure 3:* Out-of-plane diffraction data obtained at increasing cathodic potentials and calculated (DFT) diffraction data of nickel hydride ($Ni_2H$).

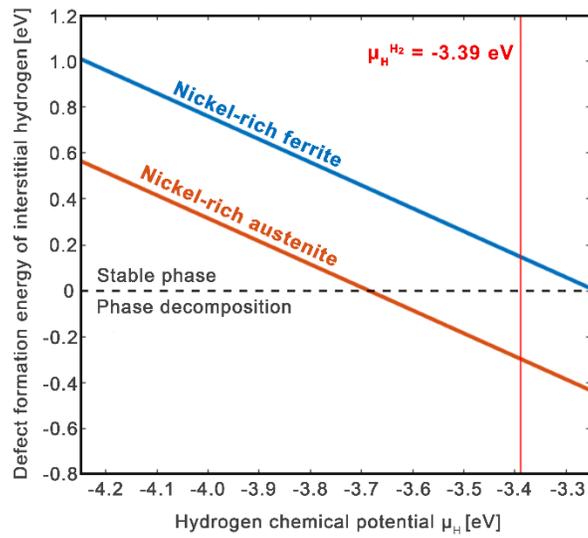

*Figure 4:* *The formation energy of interstitial hydrogen in the nickel-rich austenite phase (FCC) is significantly lower than in the ferrite phase (BCC). The dashed line demonstrates the regime of material stability for any given hydrogen chemical potential ($\mu_H$). The ferrite and austenite decompose into a new phase if the defect formation energy of interstitial hydrogen becomes negative. However, if the hydrogen chemical potential is referenced to the hydrogen molecule (-3.39 eV), only the austenite decomposes, and the ferrite remains stable.*

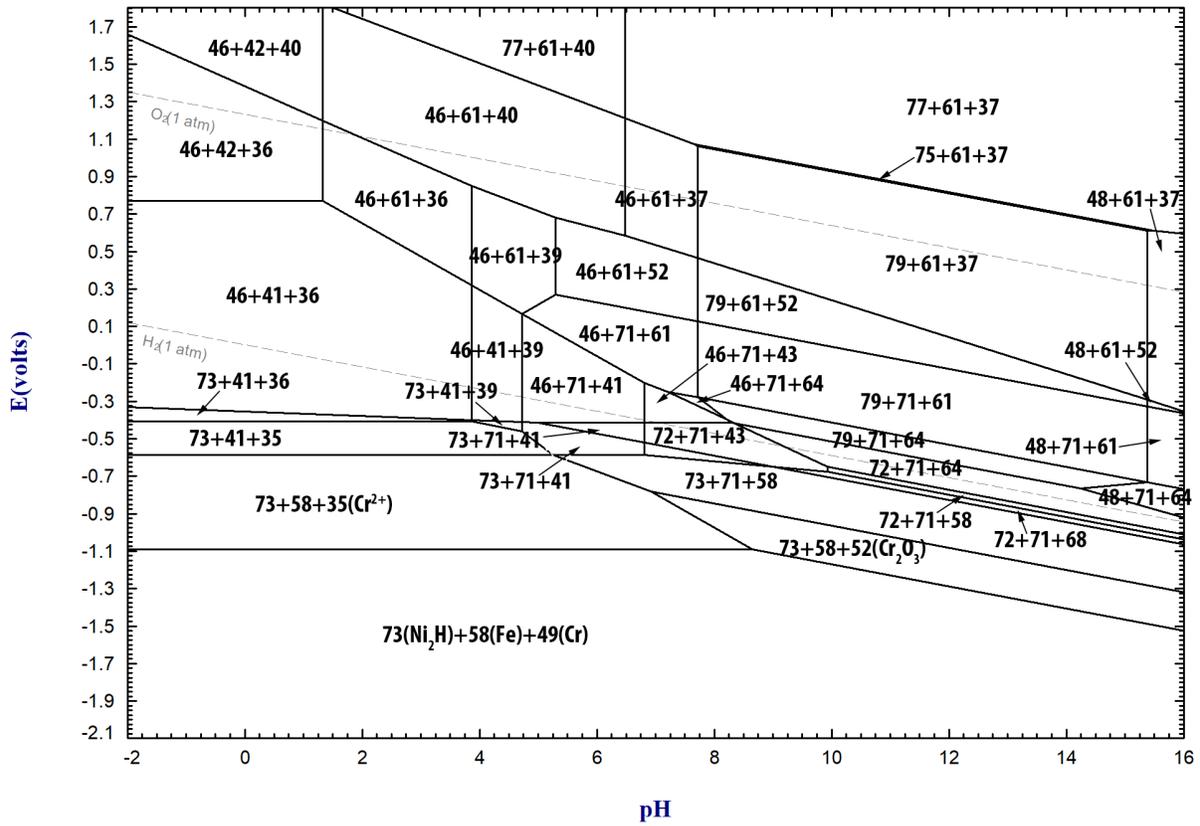

*Figure 5:* Calculated potential-pH diagram (vs standard hydrogen electrode) for the studied super duplex stainless steel. The reader is referred to **Table 2** for the labelled phases (71 is $FeCr_2O_4(s)$, 72 is Ni, 73 is $Ni_2H$). Iron and chromium are reduced to a metallic state at cathodic potentials lower than -1.1 V, where $Ni_2H$ is thermodynamically stable.

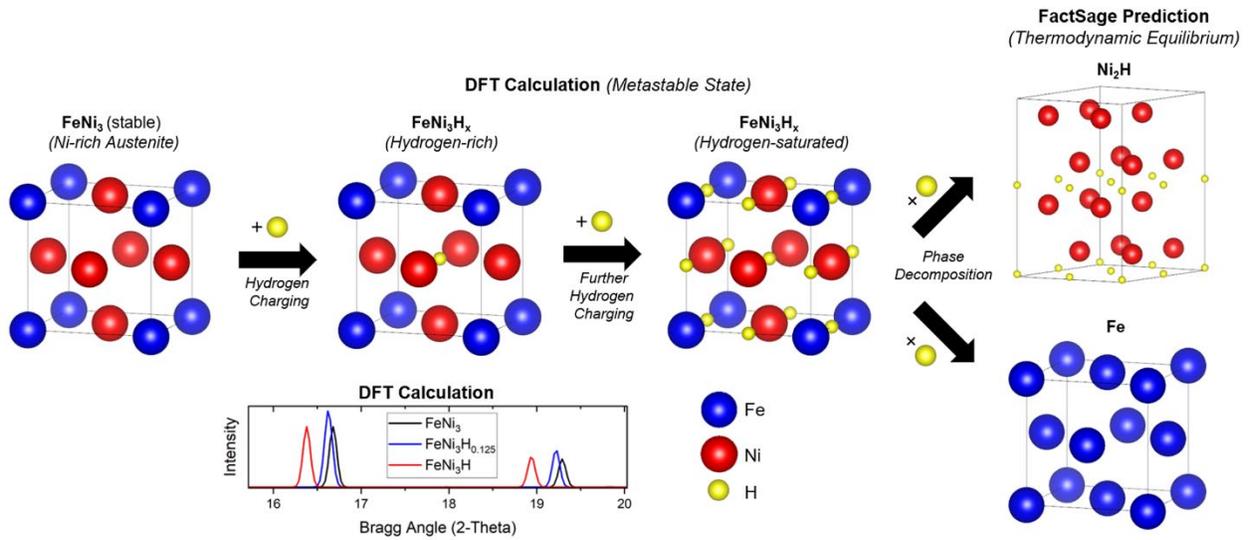

*Figure 6:* Metastable quasi-hydride formation in the nickel-rich austenite phase (surface alloy layer) of super duplex stainless steel during electrochemical hydrogen charging. The final stages are $Ni_2H$ and hydrogen-rich austenite (both stable). The entire process reverts upon the termination of cathodic polarisation.

**Supplementary**

The supercell structures of the nickel-rich ferrite, nickel-rich austenite and nickel hydride used for the DFT calculations are shown below.

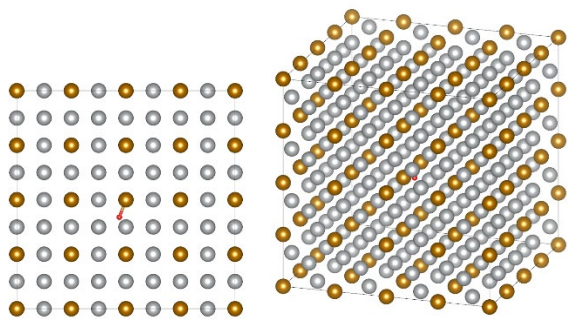

**Supplementary Figure S1:** FCC supercell ($HFe_{64}Ni_{192}$) with a hydrogen atom (red) in the cell centre with orientation projected along the vector (left) 001 and (right) 113. Iron atoms are shown in orange, and nickel atoms in grey.

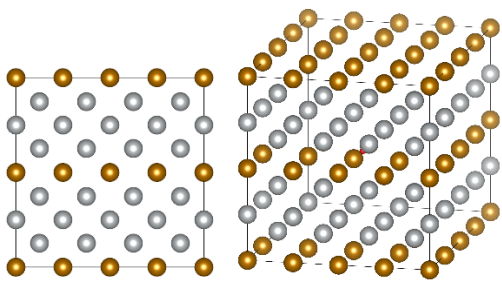

**Supplementary Figure S2:** BCC supercell ($HFe_{32}Ni_{96}$) with a hydrogen atom (red) in the cell centre with orientation projected along the vector (left) 001 and (right) 113. Iron atoms are shown in orange, and nickel atoms in grey.

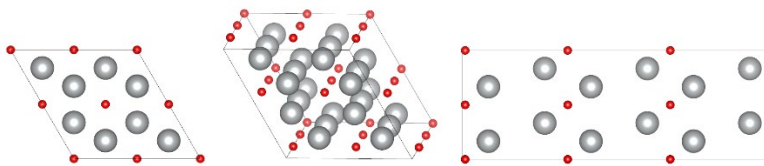

**Supplementary Figure S3:** Supercell of nickel hydride ($HNi_2$) with orientation projected along the vector (left) 001, (middle) 113, and (right) 100. Nickel atoms are shown in grey and hydrogen atoms in red.

,FactSage calculation results

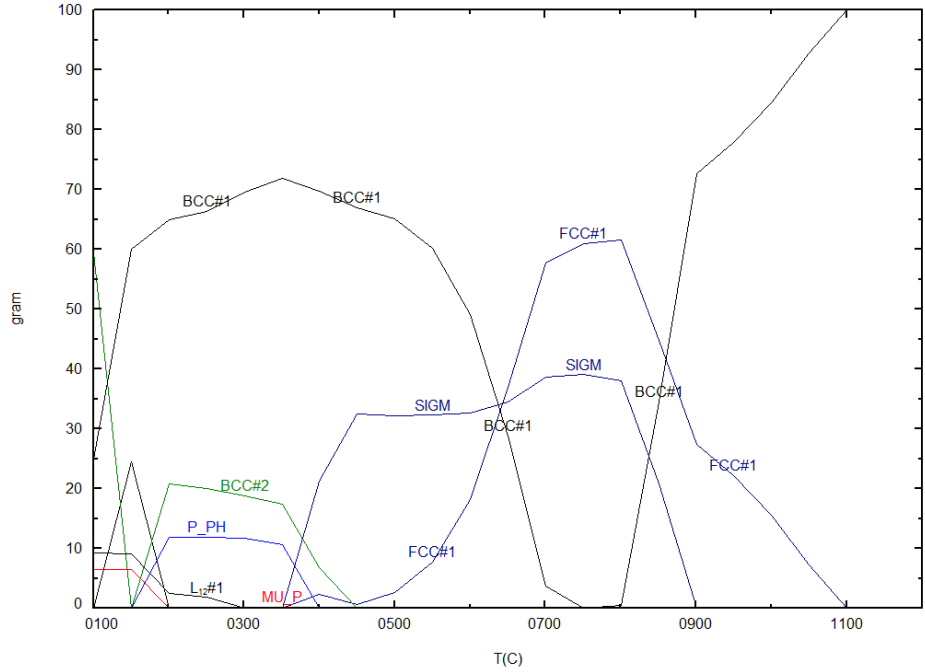

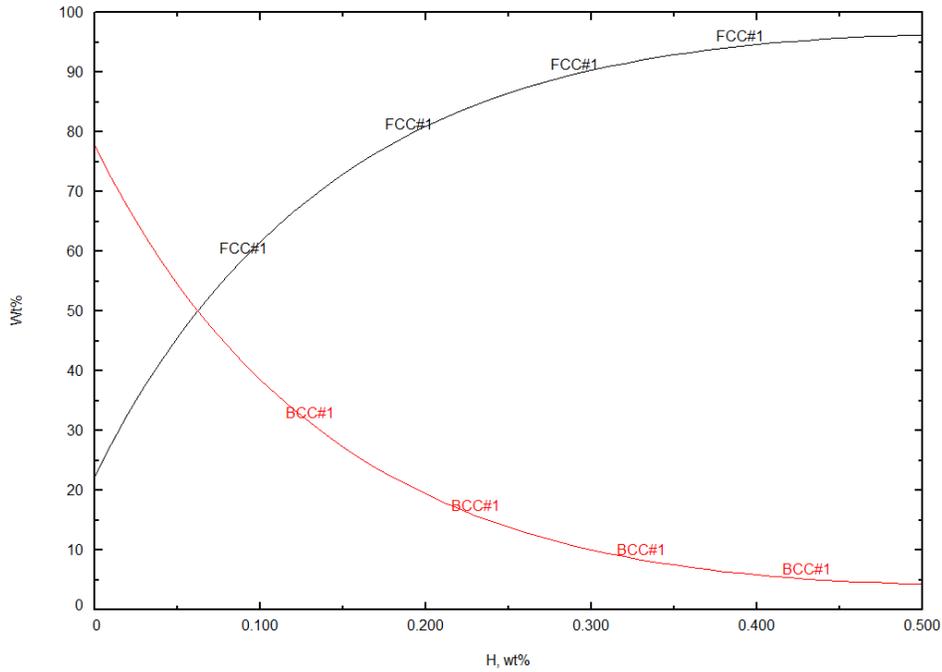

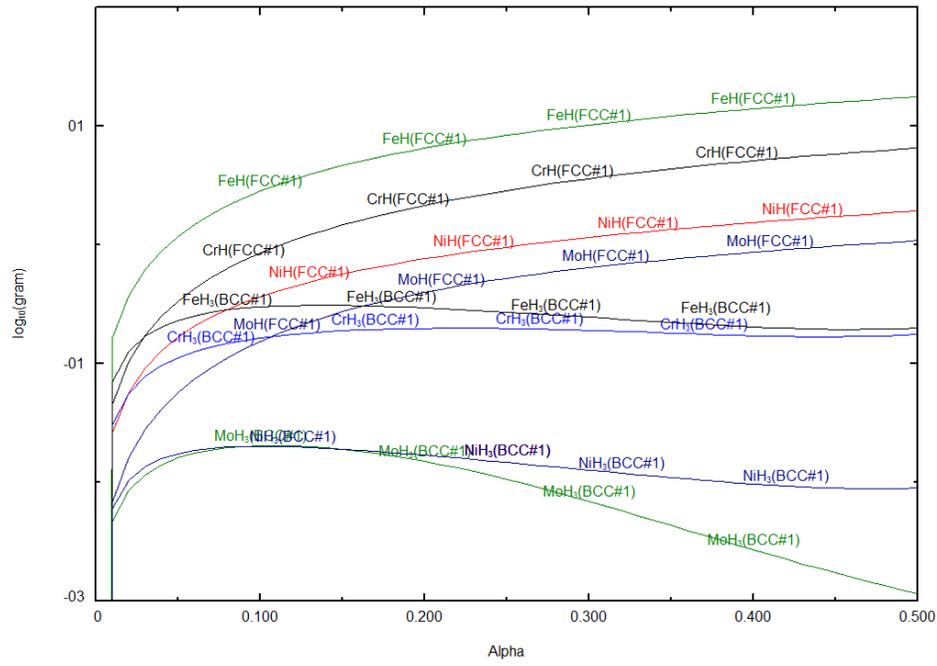

24.9 Cr + 6.9 Ni + 3.87 Mo + 64.33 Fe +

<u>At 0.5 g H addition</u>

```
  96.187    gram  FCC_A1#1
  (96.187 gram, 1.7094 mol)
+ 0        gram  FCC_A1#2
+ 0        gram  FCC_A1#3
     (950 C, 1 atm,   a=1.0000)
     ( 17.216    wt.% Cr
     + 46.596    wt.% Fe
     + 2.8937    wt.% Mo
     + 5.1021    wt.% Ni
     + 6.7682    wt.% CrH
     + 18.294    wt.% FeH
     + 1.1277    wt.% MoH
     + 2.0014    wt.% NiH)
```

Site fraction of sublattice constituents:

| | | |
|---|---|---|
| Cr | 0.25816 | Stoichiometry = 1 |
| Fe | 0.65056 | |
| Mo | 2.3512E-02 | |
| Ni | 6.7776E-02 | |

-------------------------------------------------------------

| | | |
|---|---|---|
| Va | 0.72168 | Stoichiometry = 1 |
| H  | 0.27832 | |

| System component | Amount/mol | Amount/gram | Mole fraction | Mass fraction |
|---|---|---|---|---|
| Mo | 4.0192E-02 | 3.8568 | 1.8393E-02 | 4.0097E-02 |
| Ni | 0.11586 | 6.8002 | 5.3020E-02 | 7.0698E-02 |
| Fe | 1.1121 | 62.105 | 0.50892 | 0.64566 |
| Cr | 0.44130 | 22.946 | 0.20195 | 0.23856 |
| H  | 0.47577 | 0.47955 | 0.21772 | 4.9856E-03 |

```
+ 4.3129    gram  BCC_A2#1
  (4.3129 gram, 7.9268E-02 mol)
+ 0        gram  BCC_A2#2
     (950 C, 1 atm,   a=1.0000)
     ( 41.441    wt.% Cr
     + 47.197    wt.% Fe
     + 0.27925   wt.% Mo
     + 2.1165    wt.% Ni
     + 4.0907    wt.% CrH3
     + 4.6413    wt.% FeH3
     + 2.6871E-02 wt.% MoH3
     + 0.20761   wt.% NiH3)
```

**Site fraction of sublattice constituents:**

| | | |
|---|---|---|
| Cr | 0.47409 | Stoichiometry = 1 |
| Fe | 0.50273 | |
| Mo | 1.7311E-03 | |
| Ni | 2.1450E-02 | |

-------------------------------------------------------------

| | | |
|---|---|---|
| Va | 0.91467 | Stoichiometry = 3 |
| H  | 8.5327E-02 | |

| System component | Amount/mol | Amount/gram | Mole fraction | Mass fraction |
|---|---|---|---|---|
| Mo | 1.3722E-04 | 1.3167E-02 | 1.3782E-03 | 3.0530E-03 |
| Ni | 1.7003E-03 | 9.9796E-02 | 1.7078E-02 | 2.3139E-02 |
| Fe | 3.9850E-02 | 2.2254 | 0.40027 | 0.51600 |
| Cr | 3.7580E-02 | 1.9540 | 0.37747 | 0.45307 |
| H  | 2.0291E-02 | 2.0452E-02 | 0.20381 | 4.7421E-03 |

## Single Phase austenite at 25 C

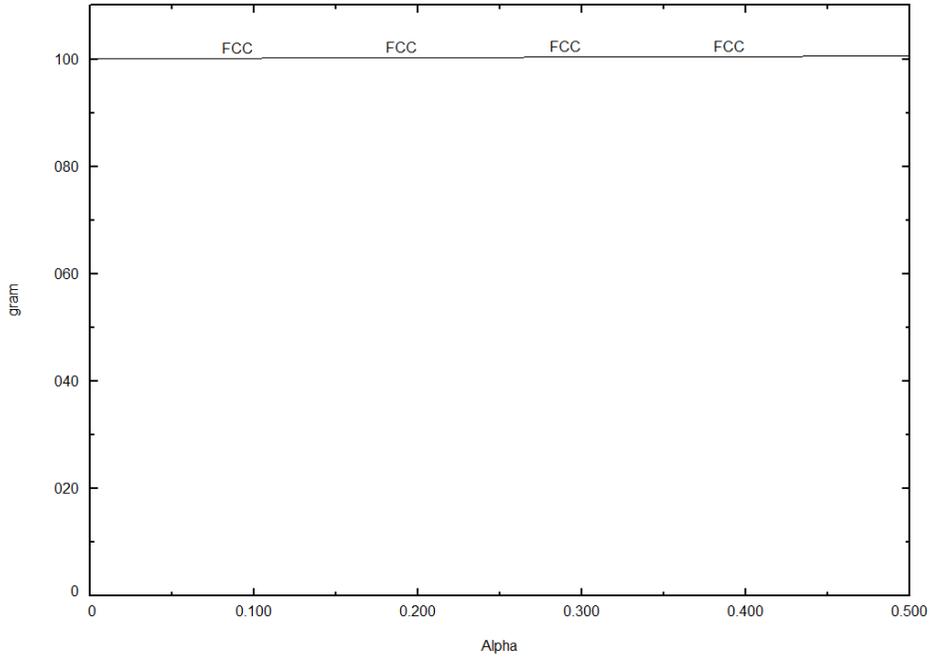

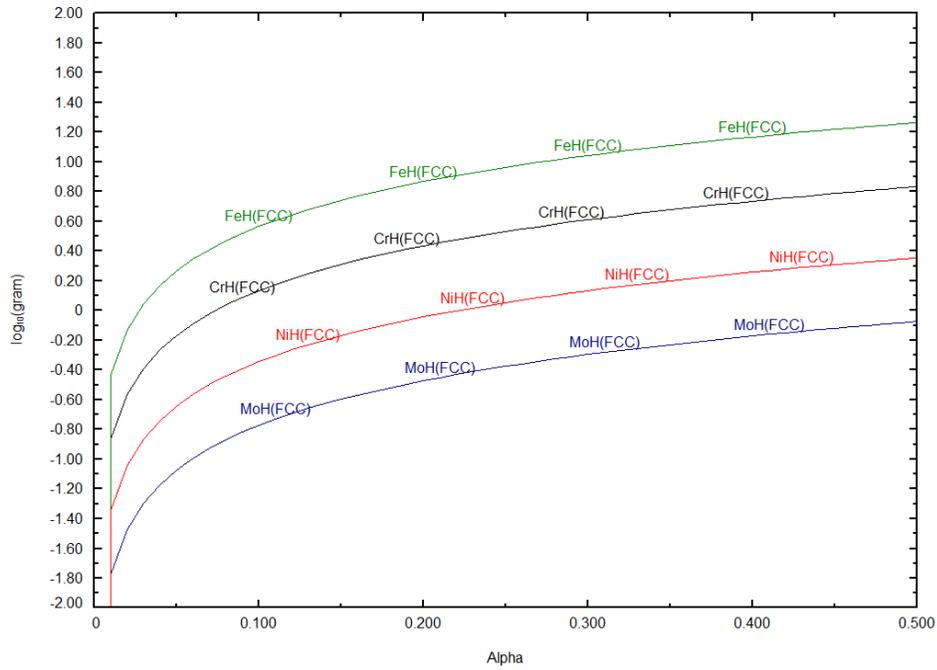

## Single Phase ferrite at 25 C

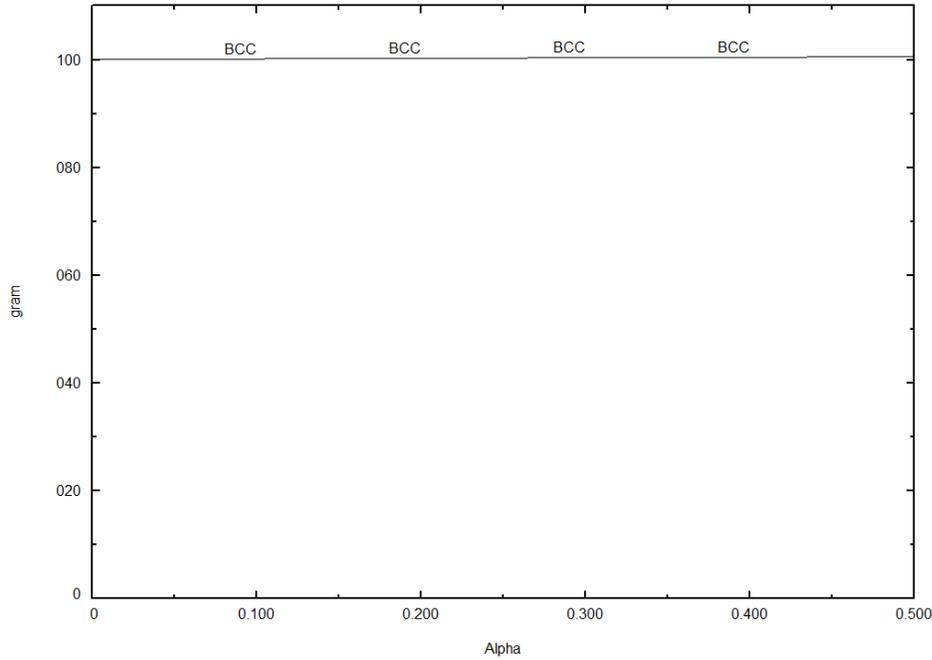

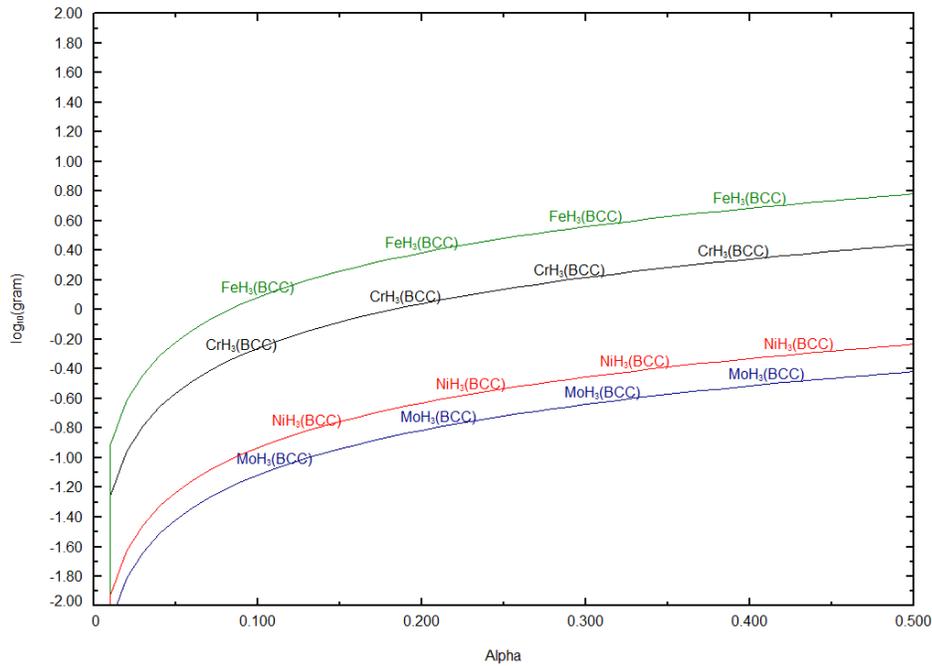

## Paraequilibrium and minimum Gibbs energy calculations

- In certain solid systems, some elements diffuse much faster than others. Hence, if an initially homogeneous single-phase system at high temperature is quenched rapidly and then held at a lower temperature, a temporary **paraequilibrium state** may result in which the rapidly diffusing elements have reached equilibrium, but the more slowly diffusing elements have remained essentially immobile.

- The best known, and most industrially important, example occurs when homogeneous austenite is quenched and annealed. Interstitial elements such as C and N are much more mobile than the metallic elements.

- At paraequilibrium, the ratios of the slowly diffusing elements in all phases are the same and are equal to their ratios in the initial single-phase alloy. The algorithm used to calculate paraequilibrium in FactSage is based upon this fact. That is, the algorithm minimizes the Gibbs energy of the system under this constraint.

- If a paraequilibrium calculation is performed specifying that no elements diffuse quickly, then the ratios of all elements are the same as in the initial homogeneous state. In other words, such a calculation will simply yield the single homogeneous phase with the minimum Gibbs energy at the temperature of the calculation. Such a calculation may be of practical interest in **physical vapour deposition** where deposition from the vapour phase is so rapid that phase separation cannot occur, resulting in a single-phase solid deposit.

- Paraequilibrium phase diagrams and minimum Gibbs energy diagrams may be calculated with the **Phase Diagram Module**. See the Phase Diagram slide show.

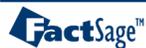 *Equilib Advanced 14.1* www.factsage.com

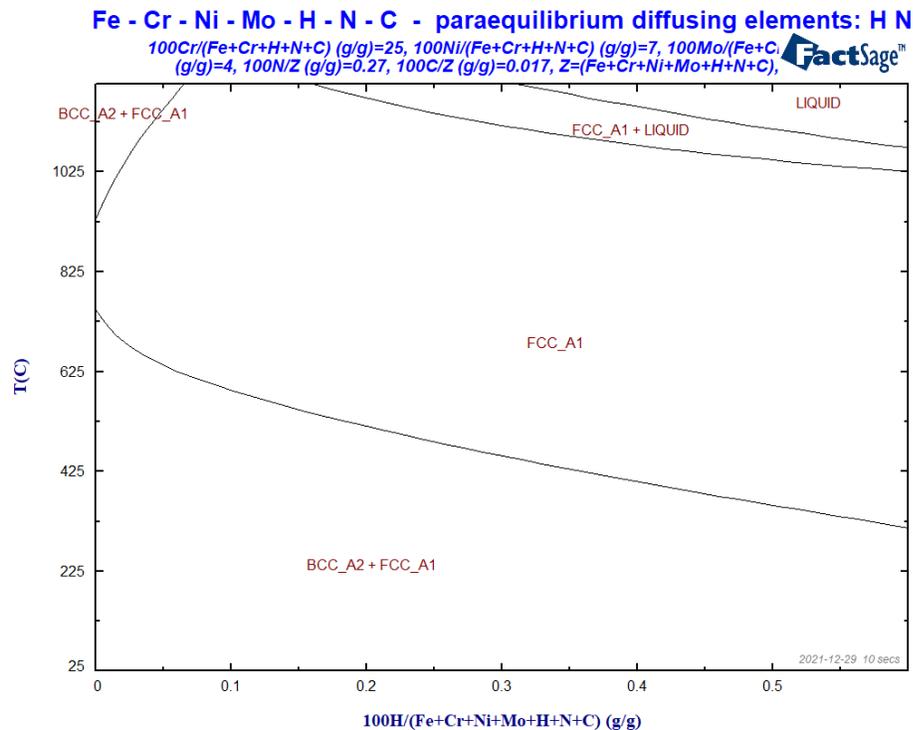

## Equilibrium phase

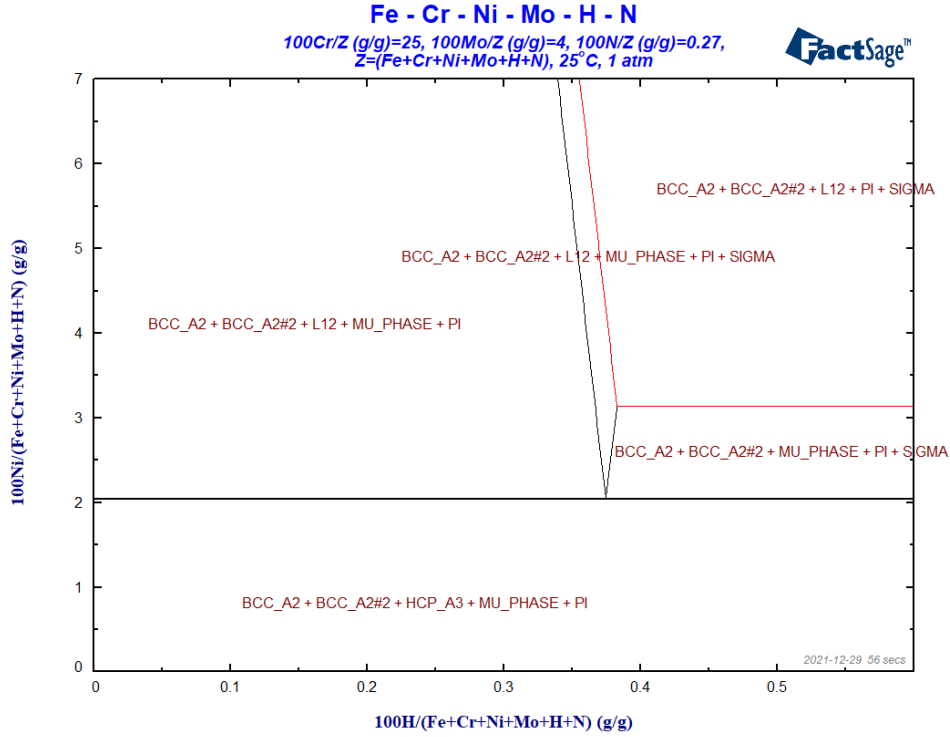

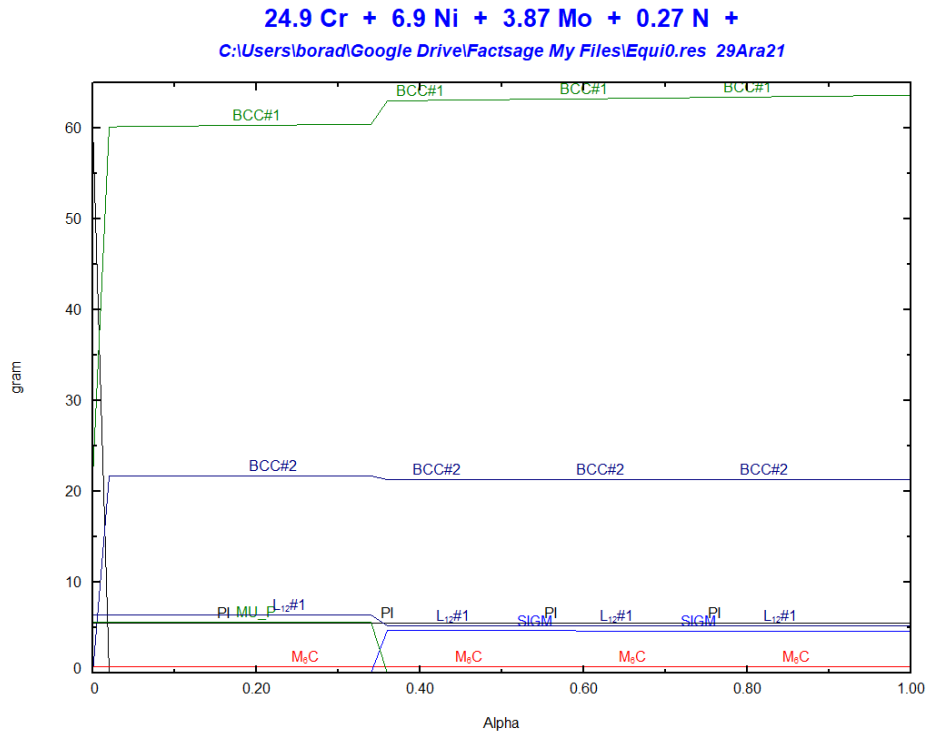

## Paraequilibrium with no diffusing element

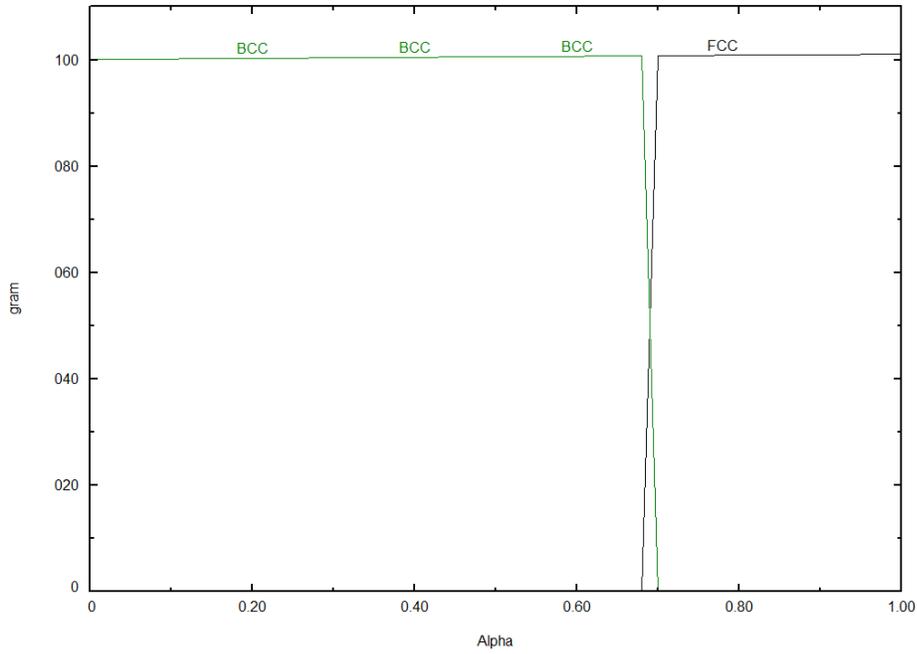

## Paraequilibrium H N diffusing element

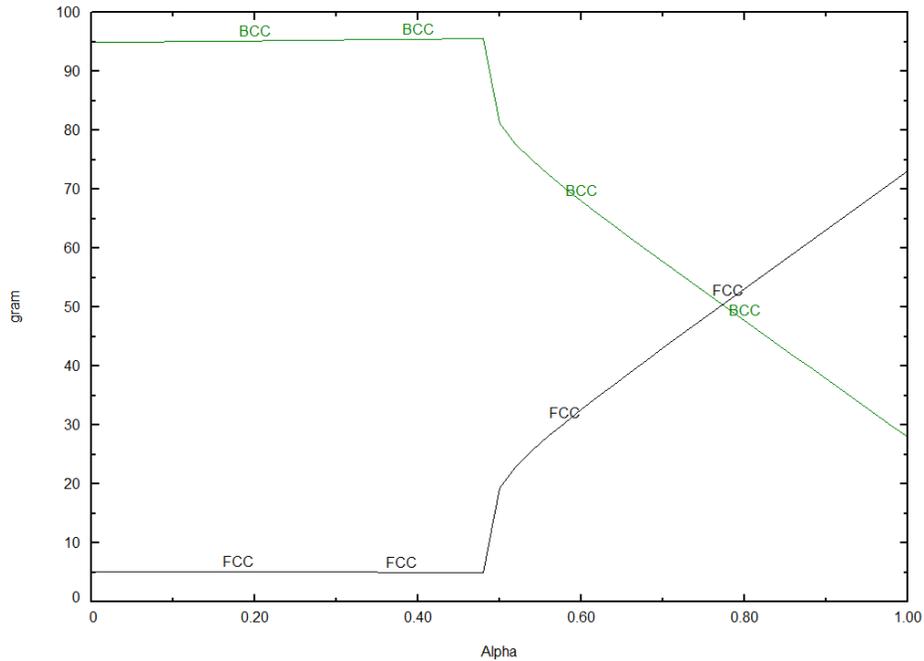

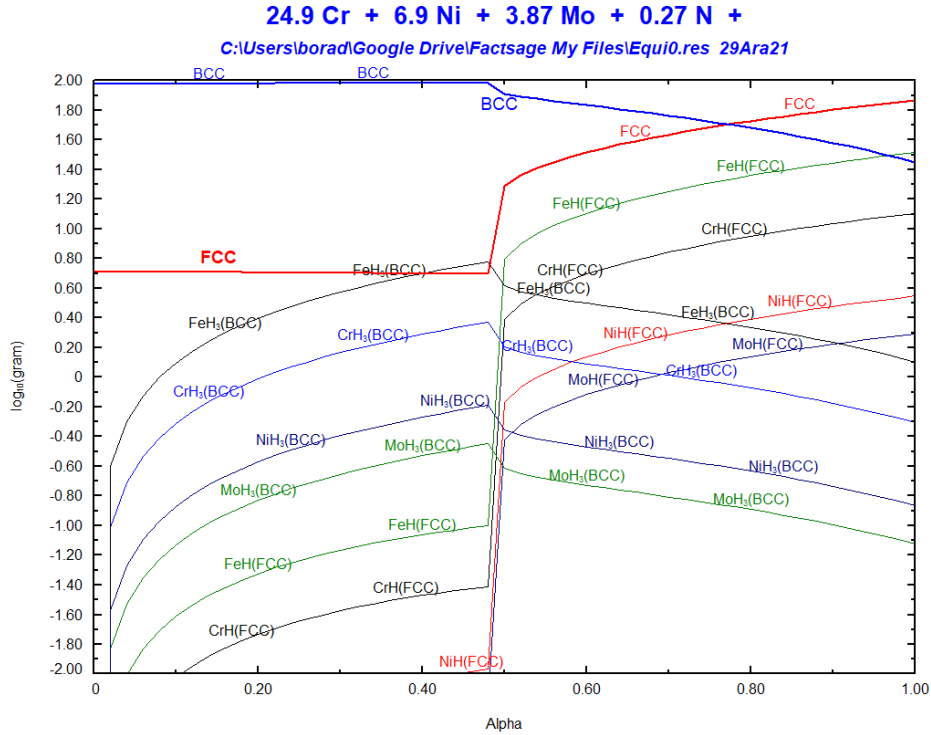

## EXCESS H in Equilibrium

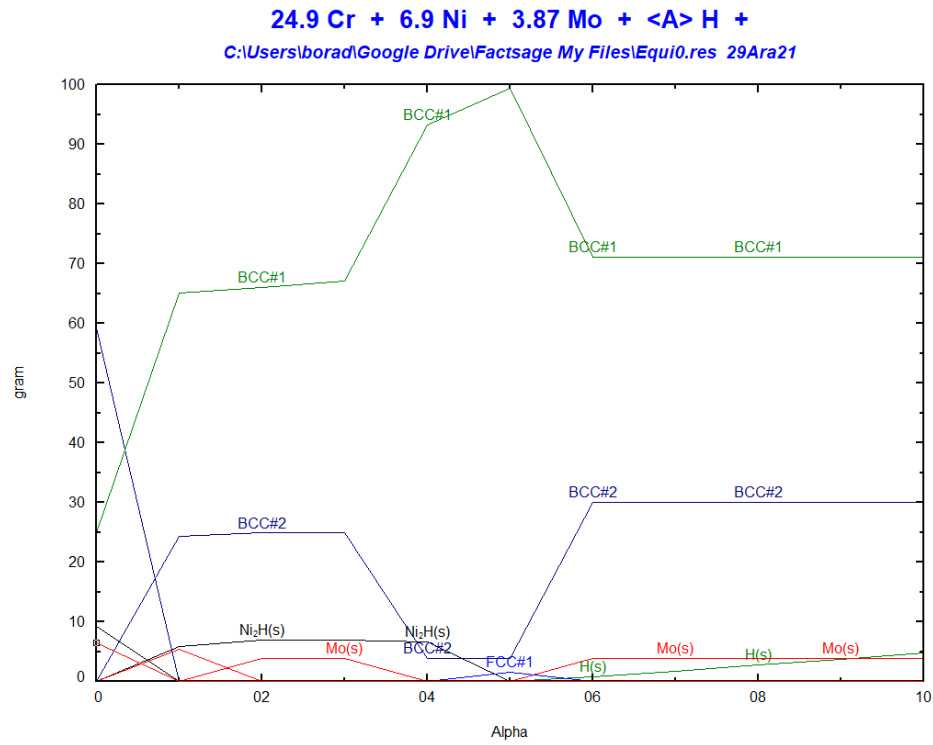

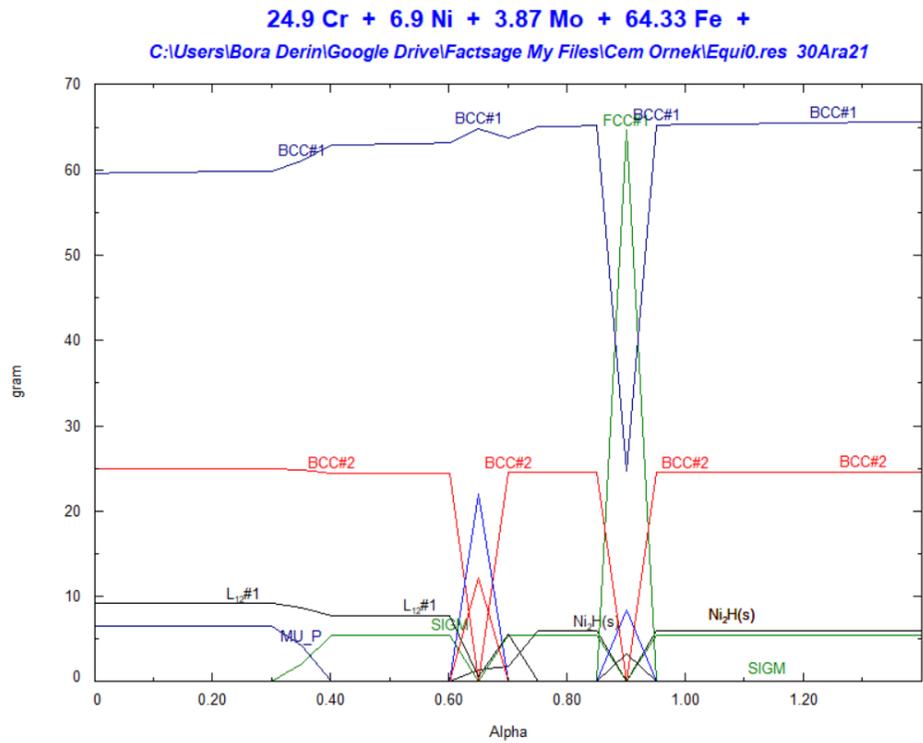